\documentclass[]{interact}

\usepackage{epstopdf}
\usepackage[caption=false]{subfig}
\usepackage{multicol}
\usepackage{optidef}
\usepackage[ruled,vlined]{algorithm2e}

\usepackage[numbers,sort&compress]{natbib}
\bibpunct[, ]{[}{]}{,}{n}{,}{,}
\renewcommand\bibfont{\fontsize{10}{12}\selectfont}

\theoremstyle{plain}
\newtheorem{theorem}{Theorem}[section]
\newtheorem{lemma}[theorem]{Lemma}

\theoremstyle{definition}

\theoremstyle{remark}
\newtheorem{remark}{Remark}

\newcommand{\diag}{{\rm diag}}

\begin{document}
\title{Study on Precoding Optimization Algorithms \\ in Massive MIMO System with Multi-Antenna Users}

 \author{
\name{Evgeny~Bobrov\textsuperscript{a,b}\thanks{CONTACT Evgeny~Bobrov. Email: eugenbobrov@ya.ru}, Dmitry~Kropotov\textsuperscript{a,c}, Sergey~Troshin\textsuperscript{c} and Danila~Zaev\textsuperscript{b}}
\affil{\textsuperscript{a}Lomonosov MSU, Russia; \textsuperscript{b}MRC, Huawei Technologies, Russia; \textsuperscript{c}HSE University, Russia}
}

\maketitle

\begin{abstract}
The paper studies the multi-user precoding problem as a non-convex optimization problem for wireless multiple input and multiple output (MIMO) systems. In our work, we approximate the target Spectral Efficiency function with a novel computationally simpler function. Then, we reduce the precoding problem to an unconstrained optimization task using a special differential projection method and solve it by the Quasi-Newton L-BFGS iterative procedure to achieve gains in capacity. We are testing the proposed approach in several scenarios generated using Quadriga~--- open-source software for generating realistic radio channel impulse response. Our method shows monotonic improvement over heuristic methods with reasonable computation time. The proposed L-BFGS optimization scheme is novel in this area and shows a significant advantage over the standard approaches. The proposed method has a simple implementation and can be a good reference for other heuristic algorithms in this field.
\end{abstract}

\begin{keywords}
Optimization, Massive MIMO, Precoding, SVD, L-BFGS, Interior-Point. 
\end{keywords}

\section{Introduction}
Wireless channels with multiple input and multiple output (MIMO) provide significantly more capacity than their counterparts with one input and one output. Thus, the MIMO system is an essential technology in modern wireless telecommunications, including Wi-Fi and 5G systems~\cite{Bjornson, bobrov2021massive}. It allows using antenna array clusters to send multiple signal beams for multiple user devices simultaneously. The proper construction of such beams is called \textit{beamforming} or \textit{precoding} procedure~\cite{Joham_RZF, RZF19}. In the linear channel assumptions, the correct precoding is a complex matrix of the general form with given constraints, which corresponds to the physical limitations of the system. We measure the quality of the obtained precoding using the well-known functions such as Signal-to-Interference-and-Noise Ratio (SINR)~\cite{wang2014sinr} and Spectral Efficiency (SE)~\cite{SE}.

The standard precoding algorithms, which are well-known in the literature, are Maximum-Ratio Transmission (MRT)~\cite{MRT}, Zero-Forcing (ZF)~\cite{ZF} and Regularized ZF (RZF)~\cite{RZF} (including its recent variant~--- Adaptive RZF (ARZF)~\cite{Conjugate}). All these algorithms have analytical formulas, without taking into account the target function of Spectral Efficiency (SE) or do it implicitly, maximizing the numerator of SINR using the MRT algorithm, or reducing the denominator of SINR using the ZF algorithm. This leads to simple but non-optimal precoding solutions. Therefore, one of our goal is to study the potential improvement of precoding methods.

While during downlink (DL) on the transmitter side a base station apply precoding matrices, symmetrically, on the receiver side users apply detection matrices. The knowledge of the detection of the user allows constructing more efficient precoding for the base station, some works even consider joint construction of precoding and detection algorithms, see e.g.~\cite{PrecodingDetection}. In general, we do not have exact information about the user detection, but we can mean a type of it. The most common detections are well-described in the literature, such as Minimum MSE (MMSE)~\cite{MMSE, MMSE2} and Interference Rejection Combiner (MMSE-IRC)~\cite{IRC}, which can be applied in order to eliminate the multiuser interference. We also utilize the recently proposed theoretical Conjugate Detection (CD)~\cite{Conjugate}.

In this article, we investigate the case when one transmitter (a base station) sends data to several receivers (users), which is also called a downlink (DL) procedure. We constrain ourselves to the case of linear transmitter and the receiver. For the case when each user has only one receiving antenna under the total power restriction at the base station, optimal linear precodings involving Mimimum MSE (MMSE)~\cite{MMSE, MMSE2} function were derived in the papers~\cite{Precoding, Precoding2, Conjugate}. In contrast, we consider the most practical formulation of the problem with multiple-antenna users~\cite{SVD} and per-antenna power constraints at the base station~\cite{Inverses}. 

Thus, considered problem is constrained optimization programming task with non-convex function and multiple constraints. It can be solved by using suboptimal heuristic methods~\cite{MRT, ZF}, interior point methods~\cite{Interior}, projection-based methods~\cite{Projection}, or any of them with reparametrization (variables transform). In this work, we find a convenient reparametrization, that reduces the constrained optimization problem to an unconstrained one. This reduction allows to use an advanced unconstrained gradient-based methods. The proper reparametrization is one of the main contributions of the article. Even with the right reparameterization, the target function is computationally complex and must be simplified to accelerate convergence. Proper simplification is the second and new approach presented in this article.

In this study we, firstly, approximate the SE~\cite{SE} objective function using the Conjugate Detection CD~\cite{Conjugate} to simplify the optimization procedure as much as possible. As the second step, we reduce the precoding problem to an unconstrained optimization task and solve using the Quasi-Newton L-BFGS iterative procedure~\cite{LBFGS} to achieve the maximum possible transmitting quality. We write an end-to-end differentiable projection-based method, where gradients are taken both for the functional and the projection, which tends to very fast convergence of the proposed method. The problem to be solved is not convex, and so we find the local maximum of the SE.

The novelty of the proposed method consists in simplifying the objective function SE, which makes the task more attractive from a computational point of view. Secondly, the reparametrization of precoding for an unconstrained task was invented. We called this parametrization \textit{Differentiable projection}. In addition, classical parameterization using the \textit{Softmax} function is also described.

Our parametrization method is a know-how and is designed to solve the problem of precoding. In fact, it can be generalized to other topics of constrained optimization problems. In combination with the simplified function, we proposed the target problem of unconstrained optimization as a reference algorithm for evaluating the upper bound of quality. Finally, all our research is based on realistic power per antenna power constrains and multiple antenna users, which in itself is a separate problem for finding the right approach to solving. 

All investigated algorithms were tested in several scenarios of Quadriga~\cite{Quadriga}~--- an open-source software for generating realistic radio channel impulse response. Our method shows monotonic improvement over heuristic methods with reasonable computational time.

We adopt commonly used notational conventions throughout the paper. Matrices and vectors will be denoted by bold-face upper and lower case letters, respectively. Furthermore, $(\cdot)^\mathrm H$ denotes Hermitian transpose.

The paper is organized as follows. In Section~\ref{sec:system_model} we describe the system model of the studied Massive MIMO network, introduce quality measures, types of detection matrices, approximated quality function, power constraints and problem statement. In Section~\ref{sec:solutions} we study solutions for precoding problem, reference methods, proposed optimization solution and its relative computational complexity. In Section~\ref{sec:experiments} we provide numerical results in the Quadriga radio simulator. Section~\ref{sec:conclusion} contains the conclusion.

\section{System Model}\label{sec:system_model}
In 5G cellular networks, we look at a precoding problem in multi-user massive MIMO communication. A single base station with multiple transmit antennas is used. These antennas send signals to multiple users at the same time. Each user also has multiple antennas for receiving signals. The transmission quality between the base station and the users is measured by the base station. The precoding problem entails determining an appropriate weighting (phase and gain) for transmitting the signal in order to maximise the signal power at the receivers' output.

\begin{figure}
    \centering
    \includegraphics[scale=0.8]{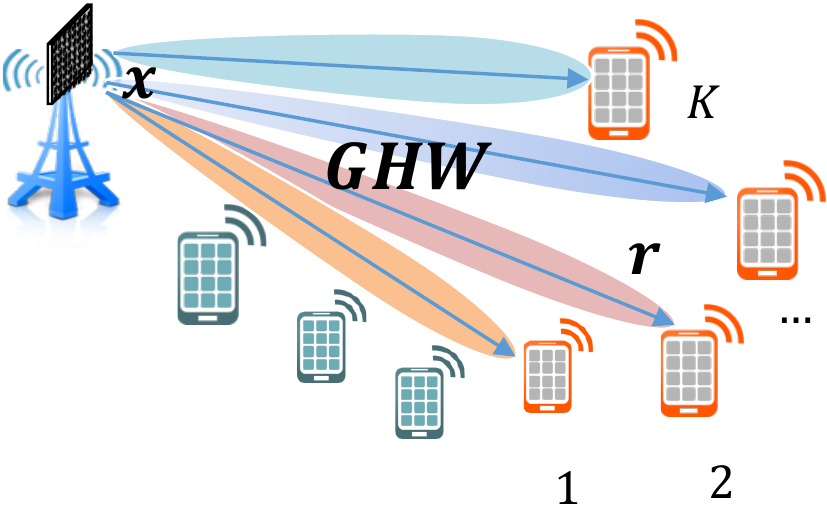}
    \caption{An example of the MIMO precoding usage.  The problem is to find an optimal precoding matrix~$\bm W$ of the system given the target SE function~\eqref{Spectral Efficiency} and the per-antenna power constraints~\eqref{eq:power_constraints}.}
    \label{fig:my_label}
\end{figure}

\begin{figure}
  \centering
  \includegraphics[width=1\linewidth]{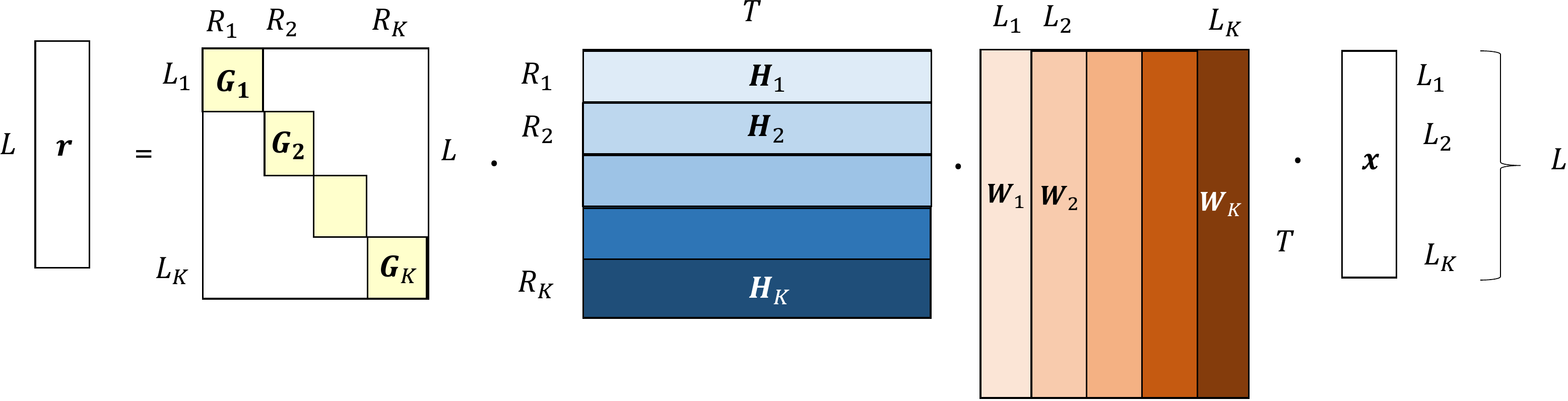}
  \caption{An example of the MIMO transmission system in the matrix form. Multi-User precoding~$\bm W$ allows transmitting different information to various users simultaneously.}
  \label{Mimo Scheme}
\end{figure}

In our system we consider $K$ \textit{users} and for $k$-th user we would like to transmit him $L_k$ symbols. In total, we would like to transmit a vector $\bm x \in \mathbb{C}^L$, where $L = L_1 + \dots + L_K$. We multiply the transmitted vector by a precoding matrix $\bm W \in \mathbb{C}^{T \times L}$, where $T$ is a total number of transmit antennas on a base station. Then we transmit the precoded symbols for the all users. Suppose that $k$-th user has $R_k$ receive antennas and $\bm H_k \in \mathbb{C}^{R_k \times T}$ is a channel between $k$-th user and the base station. Then $k$-th user receives $\bm H_k \bm W \bm x + \bm n_k$, where $\bm n_k$ is a Gaussian noise. The noise appears as a result of thermal distortions of the system. Finally, $k$-th user applies a detection of transmitted symbols by multiplying the received vector by a detection matrix $\bm G_k \in \mathbb{C}^{L_k \times R_k}$. The final vector of detected symbols of the all users is denoted by $\bm r \in C^L$. The whole process of transmitting symbols is presented in Fig.~\ref{Mimo Scheme}~\cite{Conjugate}. Note that the linear precoding and detection can be implemented by simple matrix multiplications. Usually the total number of base station antennas, user antennas and transmitted symbols are related as $L_k \leq R_k \leq T$.

Finally, the Multi-User MIMO model is described using the following linear system:
\begin{equation*}
    \bm r = \bm G ( \bm H \bm W \bm x + \bm n ).
    \label{Basic received vector}
\end{equation*}
Where $\bm r \in \mathbb{C}^L $ is a \textit{received vector}, and $\bm x \in \mathbb{C}^L$ is a \textit{sent vector}, and $\bm H \in \mathbb{C}^{R \times T}$ is a \textit{channel matrix}, and $\bm W \in \mathbb{C}^{T \times L}$ is a \textit{precoding matrix}, and $\bm G \in \mathbb{C}^{L \times R}$ is a block-diagonal \textit{detection matrix}, and $\bm n \sim \mathcal{CN}(0, I_{L}) $ is a \textit{noise-vector}.  The constant $T$ is the number of transmit antennas, $R$ is the total number of receive antennas, and $L$ is the total number of transmitted symbols in the system. They are typically related as $L \le R \le T $. Each of the matrices $\bm G, \bm H, \bm W$ decomposes by $K$ \textit{users}. The numbers $K, L, R, T$ and channel matrix $\bm H \in \mathbb{C}^{R \times T}$ are predefined in the system.

It is convenient~\cite{SVD} to represent the user channel matrix $k$ via its reduced Singular Value Decomposition (SVD):
\begin{equation*}
    \label{User k SVD}
    \bm H_k = \bm U^\mathrm H_k \bm S_k \bm V_k, \quad \bm U_k \bm U_k^\mathrm H = \bm U_k^\mathrm H \bm U_k = \bm I_{R_k}, \quad \bm S_k = \diag\{s_1,\dots,s_{R_k}\}, \quad \bm V_k \bm V^\mathrm H_k = \bm I_{R_k}.
\end{equation*}
Where the \textit{channel matrix} for user $k$, $\bm H_k \in \mathbb{C}^{R_k \times T}$ contains channel vectors $\bm h_i \in \mathbb{C}^{T}$ by rows, the singular values $\bm S_k \in \mathbb{C}^{R_k \times R_k}$ are sorted by descending, $\bm U_k \in \mathbb{C}^{R_k \times R_k}$ is a unitary matrix of left singular vectors, and matrix $\bm V_k \in \mathbb{C}^{R_k \times T}$ consists of \textit{right singular vectors}.

Collecting all users together, we can write the following decomposition:
\begin{lemma}{\label{Main Decomposition}}
\cite{Conjugate} Denote $\bm H = [\bm H_1,\dots,\bm H_K] \in \mathbb C^{R\times T}$ the concatenation of individual channel rows $\bm H_k$. The following representation holds:
\begin{equation*}    \label{User SVD}
    \bm H = \bm U^\mathrm H \bm S \bm V. 
\end{equation*}
where the $\bm H \in \mathbb{C}^{R \times T}$, 
and $\bm S = \text{diag}(\bm S_k)\in \mathbb{C}^{R \times R}$,
and $\bm U = \text{bdiag} (\bm U_k) \in \mathbb{C}^{R \times R}$ is a \textit{block-diagonal unitary matrix},
but $\bm V = [\bm V_1,\dots,\bm V_K] \in \mathbb{C}^{R \times T}$  has a general form.

\end{lemma}
\subsection{Quality Measures}

The quality functions are based not on the actual sending symbols $x \in \mathbb C^L$, but on some distribution of them~\cite{Bjornson}. Thus, we get the common functions for all the assumed symbols, which can be sent using the specified precoding. The \textit{Signal-to-Interference-and-Noise} (SINR) functional of the $l = l(k)$-th symbol and user k, is defined as:
\begin{equation}\label{Symbol SINR}
    \textrm{SINR}_l(\bm W, \bm H_k, \bm g_l, \sigma, P) = \dfrac{|\bm g_l \bm H_k \bm w_l |^2}{\sum_{i \ne l}^{L} | \bm g_l \bm H_k \bm w_i |^2 + \| \bm g_l\|^2 \frac{\sigma^2}{P}}, \quad \forall l \in \mathcal{L}_k,
\end{equation}
where $\bm w^l \in \mathbb C ^ T $ is the $l$-th column of the precoding matrix.

To get the formula for the $k$-th user SINR, namely the effective SINR, that shows the signal quality of the user, we average his $L_k$ per-symbol SINRs~\eqref{Symbol SINR} by the geometric mean:
\begin{equation}
    \textrm{SINR}_k^{eff}(\bm W, \bm H_k, \bm G_k, \sigma, P) = \Big({\prod\nolimits_{l \in \mathcal{L}_k} \textrm{SINR}_l(\bm W, \bm H_k, \bm g_l, \sigma, P) } \Big)^{\frac{1}{L_k}},
    \label{User SINR}
\end{equation}
where the detection matrix $G_k \in \mathbb C^{L_k \times R_k}$ contains vector of the $k$-th user symbols $g_l$.

To get the final functional of \textit{Spectral Efficiency}, we apply the Shannon’s formula over all effective user SINRs~\eqref{User SINR}:
\begin{equation}
    \textrm{SE}(\bm W, \bm H, \bm G, \sigma, P) =  \sum_{k=1}^K L_k \log_2 (1 + \textrm{SINR}_k^{eff}(\bm W, \bm H_k, \bm G_k, \sigma, P)) \rightarrow \max\limits_{\bm W}
    \label{Spectral Efficiency}
\end{equation}
The formula~\eqref{Spectral Efficiency} shows the throughput capacity of the system as a whole. Now it depends on the whole channel and detection matrices $\bm H \in \mathbb C^{R \times T}$ and $\bm G \in \mathbb C^{L \times R}$. We will use it as our final score function, compare algorithms and find precoding matrices $\bm W \in \mathbb C^{T \times L}$, solving the problem of maximizing this function.

Additionally, we consider the function of Single-User SINR (SUSINR):
\begin{equation}\label{SU SINR}
    \textrm{SUSINR}(\widetilde{\bm S}, \sigma^2, P) = \frac{P}{ \sigma^2} \bigg(\prod_{k=1}^K \frac{1}{L_k} \bigg(\prod\nolimits_{l \in \mathcal{L}_k} s_l^2 \bigg)^{\frac{1}{L_k}}\bigg)^{\frac 1 K}.
\end{equation}
The formula~\eqref{SU SINR} reflects the quality of the channel for the specified user without taking into account the others. It depends on the greatest $L_k$ singular values $\widetilde{\bm S}_k \in \mathbb R ^ {L_k \times L_k}$ of the $k$-th user channel matrix $\bm H_k \in \mathbb{C}^{R_k \times T}$. Finally, the matrix $\widetilde{ \bm S} \in \mathbb C^{L \times L}$ contains all singular values on the main diagonal, such as $\widetilde{ \bm S} = \text{diag} (\widetilde{ \bm S}_1 \dots \widetilde{\bm S}_K) \in \mathbb C^{L \times L}$. We will use this function as a universal channel characteristic in our experiments.

\subsection{Detection Matrices}

\textit{Detection} aims to reverse the channel with precoding to an identity matrix. We study the most common forms of the \textit{detection} matrix $\bm G$, which is called Minimum MSE (MMSE)~\cite{MMSE}, Interference Rejection Combine (MMSE-IRC)~\cite{IRC}, and Conjugate Detection (\textit{Conjugate}) (CD). In our work we will focus our attention on the MMSE-IRC and CD methods.

Firstly, MMSE detection realizes the following rule:
\begin{equation*}
    \bm G^{\textrm{MMSE}}_k(\bm A_k) = \bm A_k^\mathrm H \left(\bm A_k \bm A_k^\mathrm H + \lambda \bm I\right)^{-1}, \quad \bm A_k = \bm H_k \bm W_k.
\end{equation*}
The scalar \( \lambda = \frac{\sigma^2}{P} \) is the system noise-to-signal ratio. The constant $P$ refers to the base station power and $\sigma^2$ refers to the system noise. Thus, the method implicitly consider the \textit{noise} equal for all symbols: $\bm n = \frac{\sigma^2}{P} \bm 1 \in \mathbb C ^ L$, and detection $\bm G$ which tends to eliminate it. One may notice that the use of the method requires solving a system of linear equations of the size of layers $L$.

Secondly, MMSE-IRC detection uses additional information about noise-covariance matrix $\bm R_{uu}^{k}$ as follows:
\begin{equation}\label{IRC}
    \bm G^{\textrm{IRC}}_k(\bm A_k, \bm R_{uu}^{k}) = \bm A_k^\mathrm H (\bm A_k \bm A_k^\mathrm H + \bm R_{uu}^{k}  + \lambda \bm I)^{-1}, \quad \bm A_k = \bm H_k \bm W_k,
\end{equation}
where the covariance matrix has the form as:
\begin{equation*}
\bm R_{uu}^{k} = \bm H_{k} \left(\bm W \bm W^\mathrm H-\bm W_{k}\bm W_{k}^\mathrm H \right) \bm H_{k}^\mathrm H=\bm H_{k} \left(  \sum _{u=1,u \neq k}^{K}\bm W_{u}\bm W_{u}^\mathrm H \right) \bm H_{k}^\mathrm H.
\end{equation*}
\begin{lemma}
Using the linear property, you can get the following representation of the detection matrix MMSE-IRC:
\begin{multline*}
    \bm G^{\textit{IRC}}_k(\bm H_k, \bm R_{uu}^{k}) = \left( \bm H_{k}\bm W_{k} \right)^\mathrm H (\bm H_k \bm W_k  (\bm H_k \bm W_k)^\mathrm H + \bm R_{uu}^{k}  + \lambda \bm I)^{-1} = \\ = \left( \bm H_{k}\bm W_{k} \right)^\mathrm H (\bm H_k \bm W_k  (\bm H_k \bm W_k)^\mathrm H +  \bm H_{k} \left(  \sum _{u=1,u \neq k}^{K}\bm W_{u}\bm W_{u}^\mathrm H \right) \bm H_{k}^\mathrm H  + \lambda \bm I)^{-1} = \\ = \left( \bm H_{k}\bm W_{k} \right)^\mathrm H (\bm H_k \bm W (\bm H_k \bm W)^\mathrm H + \lambda \bm I)^{-1}.
\end{multline*}
\end{lemma}
Thirdly, \textit{Conjugate Detection} can be written in the following form: 
\begin{equation}\label{Conjugate Detection}
        \bm G^C := \widetilde{\bm S}^{-1} \widetilde{\bm U} \in \mathbb{C}^{L \times R}, \quad \bm G^C_k = \widetilde{\bm S}_k^{-1}  \widetilde{\bm U}_k \in \mathbb C ^ {L_k \times R_k}.
\end{equation}
Where $\widetilde{\bm S}_k \in \mathbb{C}^{L_k \times L_k}$ contains the first $L_k$ largest singular values, and $\widetilde{\bm U}_k \in \mathbb C ^{L_k \times R_k} $ contains the first $L_k$ corresponding singular vectors. The Conjugate and MMSE-IRC detection matrices are very closely related to each other, assuming low noise and ZF or RZF Precoding, we provide a proof of this fact in the work ~\cite{bobrov2022power}. We can use the Conjugate detection at the base station to configure the precoding matrix assuming MMSE-IRC for user equipments.

\subsection{Approximated Quality Function}

Assuming the conjugated detection $\bm G^C$~\eqref{Conjugate Detection} we have come to the following approximated function of the \textrm{SINR}~\cite{Conjugate}:
\begin{equation*}
\textrm{SINR}_l^{C}(\bm W, \widetilde{\bm v_l}, s_l, \sigma^2, P) =  \frac{ | \widetilde{\bm v_l} \bm w_l |^2 }{\sum_{i \ne l}^L| \widetilde{\bm v_l} \bm w_i |^2  +  s_l^{-2} \frac{\sigma^2}{P}}
\end{equation*}

Spectral Efficiency function can be simplified in the following way, where $\widetilde{\bm v_l} \in \mathbb C^T$ is the singular vector of the $l$-th symbol, and $\bm s_l \in \mathbb R $ is the singular value of the $l$-th symbol:
\begin{multline}
    \textrm{SE}^C(\bm W, \widetilde{\bm V}, \widetilde{\bm S}, \sigma, P) =  \sum_{l=1}^L \log_2 (1 + \textrm{SINR}_l^C(\bm W, \widetilde{\bm v_l}, s_l, \sigma, P)) = \\ 
    \sum_{l=1}^L \log_2 \Big(\sum_{i = 1}^L|\widetilde{\bm v_l} \bm w_i|^2  +  s_l^{-2} \frac{\sigma^2}{P}\Big) - \sum_{l=1}^L \log_2 \Big(\sum_{i \ne l}^L|\widetilde{\bm v_l} \bm w_i|^2  +  s_l^{-2} \frac{\sigma^2}{P}\Big) \rightarrow \max\limits_{\bm W}.
    \label{Conj Spectral Efficiency}
\end{multline}

One may notice that we have completely moved away from \textit{user antennas} of shapes $R_k$ and $R$ and work only with \textit{user layers} of shapes $L_k$ and $L$. The formula now does not depend on channel matrix $\bm H_k \in \mathbb C ^ {R_k \times T}$ but on the eigenvectors of the $l$-th layer $\bm v_l \in \mathbb C ^ T$, which has length $T$ by the number of antennas. The inversed squared singular values $s_l^{-2} \in \mathbb R$ scale the noise power.

\subsection{Power Constraints}

We formulate the realistic \textit{per-antenna power constraints}~\cite{Per_antenna_const}. Since we have $T$ equal transmitter antennas, the power limitation applied to each of them is $\frac{P}{T}$ . The antenna power can be described in terms of the row norms of the precoding matrix: 
\begin{equation}\label{eq:power_constraints}
   \|\bm w^m\|^2 \leq \frac{P}{T} \; \forall m=1\dots T. 
\end{equation}

It is clear that per-antenna constraints satisfies the total power: $\|\bm w^1\|^2 + \dots + \|\bm w^T\|^2 \leq P$, which is the \textit{sum-power constraint} across all transmit antennas. While analytically attractive, such a \textit{sum-power constraint} is often unrealistic in practice. In a physical implementation of a multi-antenna base station, each antenna has its own power amplifier in its analog front-end and is limited individually by the linearity of the power amplifier. Thus, a power constraint imposed on a \textit{per-antenna basis} is more realistic.

\subsection{Problem Statement}

We formulate the constrained smooth optimization problem:
\begin{maxi*}|l|
  {\bm W \in \mathbb C ^ {T \times L}}{\textrm{SE}(\bm W, \bm H, \bm G, \sigma, P)}{}{}
  \addConstraint{\|\bm w^m\|^2}{\leq \dfrac{P}{T}, \quad }{m = 1 \dotsc T}
  \label{Maxi}
\end{maxi*}
Where $\bm w^m \in \mathbb C ^ L $ in the $m$-th row of the precoding matrix.

\section{Solutions for Precoding}\label{sec:solutions}

\subsection{Reference Methods}

The \textit{precoding} matrix $\bm W$ is responsible for the beamforming from the base station to the users. The linear methods for precoding do the following. Firstly, the linear solutions obtain singular value decomposition for each user $\bm H_k = \bm U_k^\mathrm H \bm S_k \bm V_k \in \mathbb{C}^{R_k \times T}$ (Lemma~\ref{Main Decomposition}) and take the first $L_k$ singular vectors $ \widetilde{\bm V}_k \in \mathbb{C}^{L_k \times T}$ which attend to the first $L_k$ greatest singular values~\cite{SVD}. All these matrices are concatenated to the one matrix $\widetilde{\bm V} \in \mathbb{C}^{L \times T}$.

Finally, the precoding matrix is constructed from the obtained singular vectors. We describe linear methods for constructing a precoding matrix. We meet the power constraints using the scalar post-adjustment. We divide the precoding matrix on its maximal row norm as $\max \{ \|\bm w^t\| \}_{t=1 \dots T}$ and scale them on $\sqrt{\frac{P}{T}}$. Thus, the per-antenna power constraints can satisfy the following condition: $\|\bm w^m\|^2 \le \frac{P}{T} \; \forall m$.

\textit{Maximum-Ratio Transmission} is the simplest way of computing the precoding matrix. It takes the Hermitian conjugate of the channel singular vectors~\cite{MRT}: 
\begin{equation}\label{MRT}
    \bm W_{MRT} = \mu  \widetilde{\bm V}^\mathrm H \bm P \in \mathbb C^{T \times L}
\end{equation}
Where matrix $\bm P \in \mathbb R ^ {L \times L}$ denotes power allocation between symbols and constant $\mu > 0$ utilizes the power constraints. In such a way, the signal beams can be sent directly to users without considering their interaction. The Maximum-Ratio approach is preferred in noisy systems where the noise power is higher than inter-user interference.

\textit{Zero-Forcing} is the next modification of the precoding algorithm. which performs decorrelation of the symbols using inverse correlation matrix of the channel vectors~\cite{ZF}: 
\begin{equation}\label{ZF}
    \bm W_{ZF} = \mu \widetilde{\bm V}^\mathrm H (\widetilde{\bm V} \widetilde{\bm V}^\mathrm H)^{-1} \bm P \in \mathbb C ^ {T \times L}.
\end{equation}
With matrix $\bm P \in \mathbb R ^ {L \times L}$ and constant $\mu > 0$, which have the same meaning as in the previous section. Such precoding construction sends the signal beams to the users without creating any interference between them. Different from the previous method, the Zero-Forcing approach is preferred when the potential inter-user interference is higher than the noise power. There is a huge performance gain by eliminating the interference.

In the previous method, beams are sent not directly to the users but with some deviation, which actually reduces the payload. Thirdly, the following modification, Regularized Zero-Forcing, corrects the beams, which allows some inter-user interference and significantly increases the payload~\cite{RZF}: 
\begin{equation}\label{RZF}
    \bm W_{RZF} = \mu \widetilde{\bm V}^\mathrm H (\widetilde{\bm V} \widetilde{\bm V}^\mathrm H + \bm R)^{-1} \bm P \in \mathbb C ^ {T \times L}.
\end{equation} 
Where diagonal matrix $\bm R \in \mathbb R ^ {L \times L}$ denotes symbol \textit{regularization} and matrix $\bm P \in \mathbb R ^ {L \times L}$ and constant $\mu > 0$ as the previous ones. As the baseline, we use a special form~\cite{OptimalRegularization} of the regularization matrix $\bm R = \frac{\sigma^2 L }{P} \bm I$. It is the most common precoding in real practice.

Algorithm $\bm W_{RZF}$ is a precoding with \emph{scalar regularisation}. Finally, \textit{Adaptive Regularized Zero-Forcing} takes into account effective noise of the system. It was proposed Adaptive RZF (ARZF) algorithm~\cite{Conjugate} with \emph{diagonal matrix regularization}:
\begin{equation}\label{ARZF}
\bm W_{ARZF} = \mu \widetilde{\bm V}^\mathrm H(\widetilde{\bm V} \widetilde{\bm V}^\mathrm H + \lambda \bm S^{-2})^{-1} \bm P, \quad \lambda = \frac{\sigma^2 L}{P}.
\end{equation}
\subsection{Proposed Optimization Solution}

Let us introduce the following parametrization. This method explicitly constrains antenna rows using (sub-)differentiable projection on a ball as a part of the Conjugate Spectral Efficiency function:
\begin{maxi*}|l|
  {\bm W \in \mathbb C ^ {T \times L}}{\mathcal S(\text{proj}_{P, T}(\bm W))}{}{},
  \label{Conj Maxi}
\end{maxi*}
where $\mathcal S(\bm W)$ is defined assuming CD $\mathcal S(\bm W) = \textrm{SE}^C(\bm W, \widetilde{\bm V}, \bm S, \sigma, P)$ or assuming MMSE-IRC detection
$\mathcal S(\bm W) = \textrm{SE}(\bm W, \bm H, \bm G^{IRC}(\bm W), \sigma, P)$ and
\begin{equation*}
  \text{proj}_{P,T}(\bm W) =
    \begin{cases}
      \bm w^m , & \| \bm w^m \|^2 \leq \frac{P}{T}\\
      \frac{\bm w^m}{\|\bm w^m \|} \sqrt{\frac{P}{T}}, & \| \bm w^m \|^2 > \frac{P}{T}, \quad  \forall m = 1 \dotsc T 
    \end{cases}       
\end{equation*}
with $\bm w^m \in \mathbb C ^ L $ in the $m$-th row of the precoding matrix.

\begin{remark}
We also add maximization of the target SE function for reference.
\end{remark}

\begin{remark}
As the starting point, we are using Regularized $\bm W_{RZF}$ or Adaptive $\bm W_{ARZF}$.
\end{remark}

The key feature of the proposed method is that the optimization process, on the one hand, always remains within the feasible range and at the same time can be solved using unconstrained optimization methods, such as quasi-Newton methods. In contrast to the projection gradient method, the iteration solution always remains in a feasible area and does not need an explicit projection on the boundary. Moreover, in the particular case the projection-based method doesn't converge appropriately. And the same for interior point method: its basic realisation doesn't give an appropriate solution for our task. An advantage of the proposed method is that it seeks the solution on the boundary (and we know that the optimum is on the boundary). This search within the such complicated edge is possible using the Quasi-Newton L-BFGS method~\cite{LBFGS}.

\begin{algorithm}
\SetAlgoLined
\KwIn{Channel singular vectors $\widetilde{\bm V}$, \\ channel singular values $\bm S$, station power $P$, noise $\sigma^2$, iterations $N$,}
Tolerance grad $\varepsilon_g$, termination tolerance on first order optimality (default: 1e-5),
\\
Tolerance change $\varepsilon_c$, termination tolerance on function value and parameter changes (default: 1e-9).
\\
{\bf Initialize} precoding matrix $\bm W \leftarrow \bm W_0$\; 

 \For{$t=1$ \KwTo $N$}{
 \If{True conditions on $\varepsilon_g$ or $\varepsilon_c$}{
     \Return{$\text{proj}_{P, T}(\bm W)$}}
  Calculate the gradient: $\nabla_{\bm W} \mathcal S(\text{proj}_{P, T}(\bm W))$\; 
  Find the optimal direction recursively: $\bm D = \bm D (\nabla_{\bm W} \mathcal S(\text{proj}_{P, T}(\bm W)))$\;
  Find the optimal step length $\alpha = \arg\max\limits_\alpha \mathcal S(\text{proj}_{P, T}(\bm W + \alpha \bm D))$\;
  Make the optimization step: $\bm W \leftarrow \bm W + \alpha \bm D$\;
 }
\Return{$\text{proj}_{P, T}(\bm W)$}
\caption{On the optimal precoding matrix and Quasi-Newton L-BFGS~\cite{LBFGS}}
\label{Quasi-Newton}
\end{algorithm}

We consider $\mathcal S(\bm W)$ assuming CD or assuming MMSE-IRC detection, and also initial precoding matrix in the form of RZF or ARZF. By this way, we compare four options:~\begin{multicols}{2}
\begin{enumerate}
    \item QN CD RZF
    \item QN CD ARZF
    \item QN IRC RZF
    \item QN IRC ARZF
\end{enumerate}
\end{multicols}

\subsection{Relative Computational Complexity}

The complexity of our algorithm is no more than $\sim 3Nx$ times higher than the baseline RZF. On each iteration of QNS we have to compute gradient $\nabla \textrm{SE}^C(\bm W)$ of $x$ complexity and adjust step length $\alpha$, which takes roughly two calls of the function $\textrm{SE}^C(\bm W)$ of $2x$ total complexity (see the Table~\ref{Complexity}).

\begin{table}
    \centering
    \begin{tabular}{c|c|c|c}
         RZF & $\textrm{SE}^C(\bm W)$ & $\nabla \textrm{SE}^C(\bm W)$ & N-iterations of QNS  \\
        \hline
        $x$ & $x$ & $x$  & $\sim 3Nx$ \\
    \end{tabular}
\caption{Relative computational complexity.}
\label{Complexity}
\end{table}

\subsection{Alternative Optimization Solution}

For comparison, let us consider the classic softmax parametrization of the precoding matrix (this is an option of interior point approach). It turned out that such parametrization leads to the same solution as the differentiable projection method, but the latter converges at least twice as fast as softmax. We believe that this is due to the fact that most constraints should be up to the boundary, and softmax parameterization arguments in this sense should have the value of infinity. So, we marked the softmax method as an alternative method and brought it here for a better understanding of the solvable problem.
\begin{maxi*}|l|
  {\bm W \in \mathbb C ^ {T \times L}}{\textrm{SE}^C(\bm W, \widetilde{\bm V}, \bm S, \sigma, P)}{}{}
  \label{Sofmax Maxi}
\end{maxi*}
\begin{equation*}
    \bm W = \{w_{ij}\}_{i,j=1}^{T, L}, \quad w_{ij} = \rho_{ij} \exp(\bm i \phi_{ij}), \quad \rho_{ij} \in \mathbb R_+, \quad \phi_{ij} \in [0, 2\pi)
\end{equation*}
\begin{equation*}
    \rho_{ij}^2 = \frac{\exp(\theta_{ij})}{\sum_{k=1}^{L}\exp(\theta_{ik})}\frac{ \sigma(\alpha_i)P}{T}, \quad \phi_{ij} = 2\pi \sigma (\eta_{ij})
\end{equation*}
Where $\sigma(\cdot)$ means a sigmoid function, and $\{\theta_{i j}, \eta_{ij}, \alpha_i \}$ are free parameters that can take any real values. Thus, in another way, we can reduce the problem of optimizing precoding with constraints to the problem without constraints.

\section{Numerical Experiments}\label{sec:experiments}

\begin{figure}
    \centering
    \includegraphics[width=0.8\linewidth]{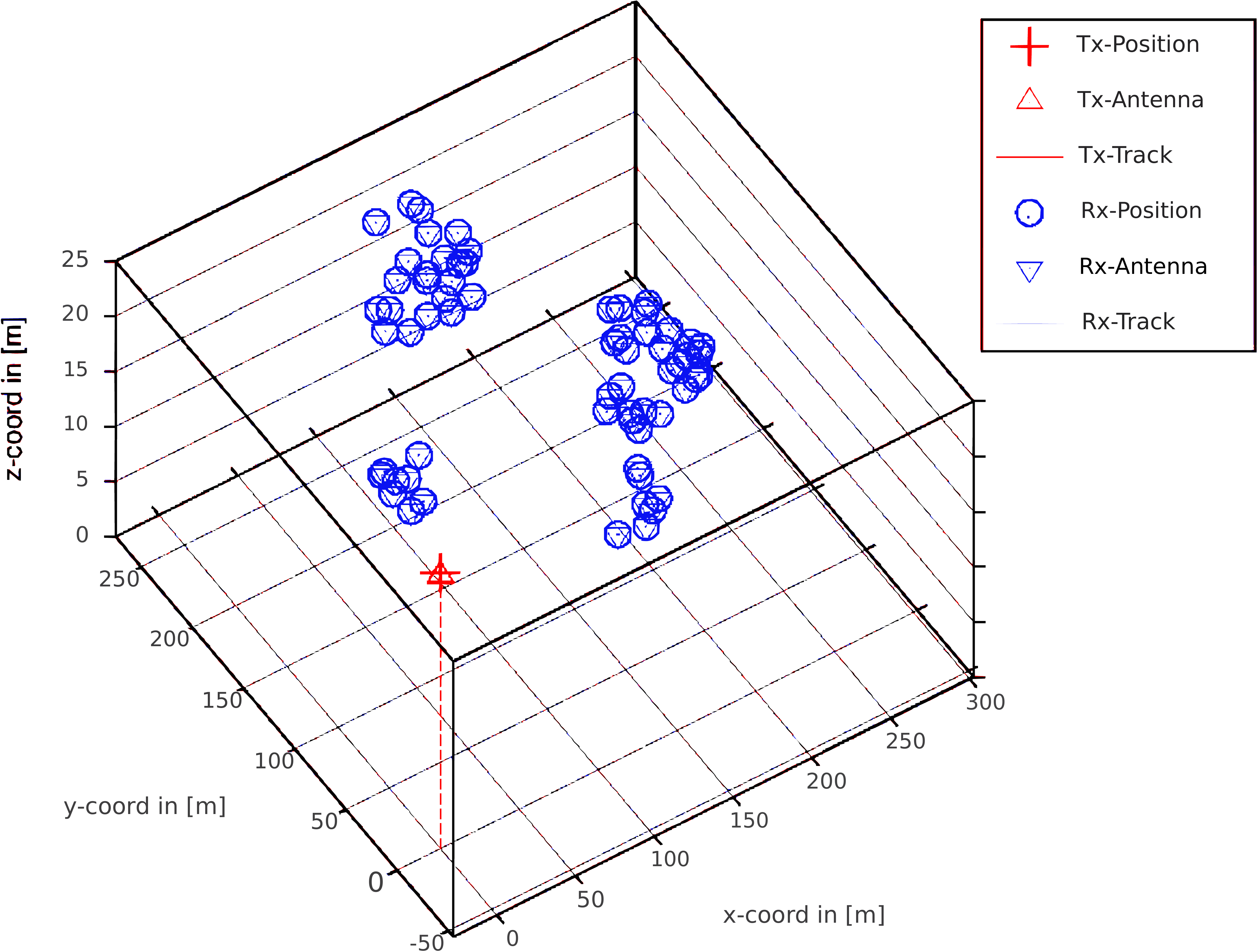}
    \caption{Example of Users for Urban Two Building Setup~\cite{Conjugate}.}
    \label{Two Buildings}
\end{figure}

The main scenario is Urban Non-Line-of-Sight 3GPP\_38.901\_RMa\_NLOS~\cite{LOS} (Fig.~\ref{Two Buildings}). We model users in the urban landscape. We are testing the proposed approach in several scenarios generated using Quadriga~\cite{Quadriga}~--- open-source software for generating realistic radio channel impulse response.  For each scenario, we generate 40 different channel matrices $H \in \mathbb{C}^{n_{users} \times 4 \times 64}$. The carrier frequency for each channel matrix is selected randomly over the bandwidth. User selection is described in the next section. The base station antenna array forms a grid with $8$ placeholders along the $y$ axis and $4$ placeholders along the $y$ axis. The receiver antenna array consists of two placeholders along the $x$ axis. Each placeholder contains two cross-polarized antennas. An interested reader can find detailed hyperparameters for antenna models and generation processes in table ~\ref{tab:quadriga_hyp}. All unlisted Quadriga parameters are those set by default.We describe the generation process in detail in our work~\cite{Conjugate}.

For each scenario we generate 40 different channels: $\bm H \in \mathbb{C}^{K \times R_k \times T}$:
\begin{itemize}
    \item $T = 64$ base station antennas,
    \item $K = 8$ users,
    \item $R_k = 4$ user antennas,
    \item $L_k = 2$ user layers.
\end{itemize}

\begin{figure}
    \centering
    \includegraphics[width=\linewidth]{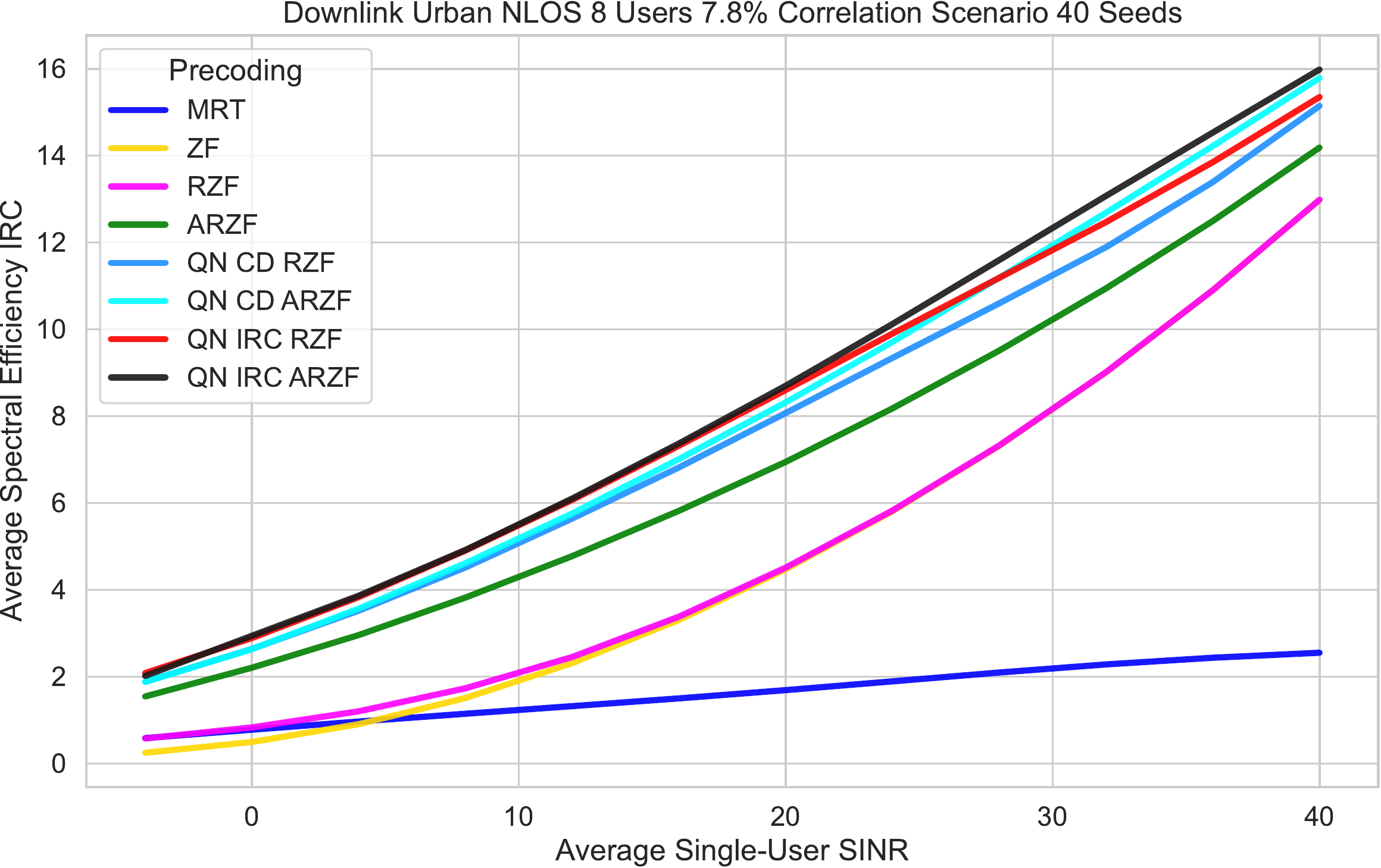}
    \caption{Urban NLOS 8 Users. The graph shows IRC Spectral Efficiency of the different precodings.}
    \label{Urban NLOS 8 Users Numerical Results}
\end{figure}

\subsection{Qualitative Results}

For each generated channel seed, we fix station power $P=1$ and select system noise $\sigma^2$ for the specified SUSINR~\eqref{SU SINR} average value.  We report the hyper-parameters for Quadriga in the Table~\ref{tab:quadriga_hyp}.

The results of the comparison are in Figure~\ref{Urban NLOS 8 Users Numerical Results} and Table~\ref{Urban NLOS 8 Users SE-IRC}. In all cases, the final SE function is calculated using MMSE-IRC detection~\eqref{IRC}, geometric mean effective SINR~\eqref{User SINR} and Shannon's capacity~\eqref{Spectral Efficiency}. 

To begin with, the experimental results in Figure~\ref{Urban NLOS 8 Users Numerical Results} and Table~\ref{Urban NLOS 8 Users SE-IRC} show that all gradient-based methods starting with "QN," which stands for "Quasi-Newton" (Algorithm~\eqref{Quasi-Newton}), outperform heuristic algorithms such as MRT~\eqref{MRT}, ZF~\eqref{ZF}, RZF~\eqref{RZF}, and ARZF~\eqref{ARZF}. 

The second keyword "CD" and "IRC" stands for the assumed detection algorithm: Conjugate Detection~\eqref{Conj Spectral Efficiency} and MMSE-IRC~\eqref{Spectral Efficiency} functions, respectively, and their gradients. It is important to observe that the performance of the obtained method is influenced by the target optimization function. Remember that the MMSE-IRC function was used to calculate the final quality measure. As a result, the "IRC" method, which optimises the target function directly, has a higher final quality measure.

\begin{figure}
    \centering
    \includegraphics[width=\linewidth]{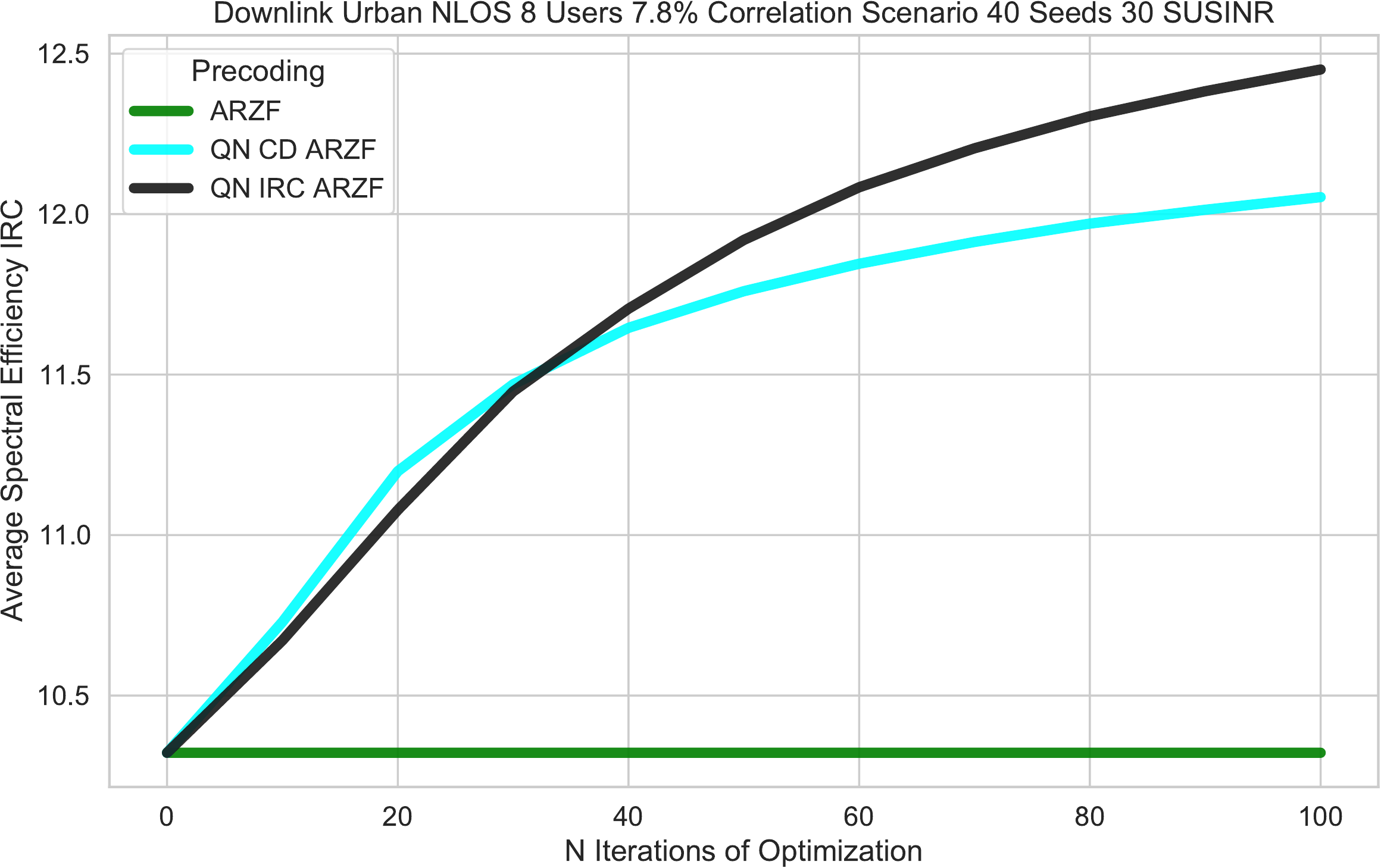}
    \caption{The graph shows IRC Spectral Efficiency on the algorithm iterations.}
    \label{Urban NLOS 8 Users Iterations}
\end{figure}

On the other hand, the "CD" method, on the other hand, converges faster through iterations and each iteration is computationally simpler than IRC.  On the Fig.~\ref{Urban NLOS 8 Users Iterations} one can find the quality of the proposed gradient methods by iterations. We noticed that modification of algorithm that uses approximated "CD" optimization function gives better results than "IRC" target function in the first iterations. At higher iterations the "IRC" gradient method superiors the "CD" method.

The third keyword in the title of gradient methods, "RZF"~\eqref{RZF}  and "ARZF"~\eqref{ARZF}, refers to the method's starting point. The method will be more convergent if the starting point is better. Methods based on the "ARZF" starting point outperform methods based on the "RZF" point in this way. Remember that our issue isn't concave. As a result, we only find the local maximum, and a good starting point is critical for our method to work.

\section{Conclusion}\label{sec:conclusion}

The paper studies the multi-user precoding problem as a non-convex optimization problem for wireless MIMO systems. In this study we approximate the Spectral Efficiency objective function using a differentiable projection-based method, where gradients are taken both for the function and the projection, which tends to very fast convergence of the proposed method. We then reduce the precoding problem to an unconstrained optimization task and solve it by the Quasi-Newton L-BFGS iterative procedure. Finally, all our research is based on realistic power limits per antenna and multiple antenna users, which in itself is a standalone problem for finding the right approach to solving. All investigated algorithms were studied in massive experiments using Quadriga software. The proposed method shows monotonic improvement over heuristic methods with reasonable computation time. It has a simple implementation and can be a good reference for other heuristic algorithms in the MIMO field. 

\section*{Acknowledges}
Authors are very grateful for D.~Minenkov for fruitful discussions. 

\section*{Funding}
The work is funded by Huawei Technologies.

\bibliographystyle{tfs}
\bibliography{interacttfssample}

\begin{thebibliography}{10}
\providecommand{\MR}{\relax\unskip\space MR }
\providecommand{\url}[1]{\normalfont{#1}}
\providecommand{\urlprefix}{Available at }

\bibitem{Precoding2}
B. Bandemer, M. Haardt, and S. Visuri, \emph{Linear MMSE multi-user MIMO
  downlink precoding for users with multiple antennas}, in \emph{2006 IEEE 17th
  International Symposium on Personal, Indoor and Mobile Radio Communications}.
  IEEE, 2006, pp. 1--5.

\bibitem{Bjornson}
E. Bj{\"o}rnson, J. Hoydis, and L. Sanguinetti, \emph{Massive mimo networks:
  Spectral, energy, and hardware efficiency}, Foundations and Trends in Signal
  Processing 11 (2017), pp. 154--655.

\bibitem{Conjugate}
E. Bobrov, B. Chinyaev, V. Kuznetsov, H. Lu, D. Minenkov, S. Troshin, D.
  Yudakov, and D. Zaev, \emph{Adaptive regularized zero-forcing beamforming in
  massive mimo with multi-antenna users}, arXiv preprint arXiv:2107.00853
  (2021).

\bibitem{bobrov2022power}
E. Bobrov, B. Chinyaev, V. Kuznetsov, D. Minenkov, and D. Yudakov, \emph{Power
  allocation algorithms for massive mimo systems with multi-antenna users}
  (2022).

\bibitem{bobrov2021massive}
E. Bobrov, D. Kropotov, and H. Lu, \emph{Massive mimo adaptive modulation and
  coding using online deep learning algorithm}, IEEE Communications Letters
  (2021).

\bibitem{LOS}
F. Bohagen, P. Orten, and G. Oien, \emph{Construction and capacity analysis of
  high-rank line-of-sight MIMO channels}, in \emph{IEEE Wireless Communications
  and Networking Conference, 2005}, Vol.~1. IEEE, 2005, pp. 432--437.

\bibitem{Projection}
J.C. Chen, \emph{Gradient projection-based alternating minimization algorithm
  for designing hybrid beamforming in millimeter-wave mimo systems}, IEEE
  Communications Letters 23 (2018), pp. 112--115.

\bibitem{Interior}
H.H. Dam and A. Cantoni, \emph{Interior point method for optimum zero-forcing
  beamforming with per-antenna power constraints and optimal step size}, Signal
  processing 106 (2015), pp. 10--14.

\bibitem{Quadriga}
S. Jaeckel, L. Raschkowski, K. B{\"o}rner, and L. Thiele, \emph{Quadriga: A 3-d
  multi-cell channel model with time evolution for enabling virtual field
  trials}, IEEE Transactions on Antennas and Propagation 62 (2014), pp.
  3242--3256.

\bibitem{Joham_RZF}
M. Joham, W. Utschick, and J.A. Nossek, \emph{Linear transmit processing in
  mimo communications systems}, IEEE Transactions on signal Processing 53
  (2005), pp. 2700--2712.

\bibitem{Precoding}
M. Joham, W. Utschick, and J.A. Nossek, \emph{Linear transmit processing in
  mimo communications systems}, IEEE Transactions on signal Processing 53
  (2005), pp. 2700--2712.

\bibitem{MRT}
T.K. Lo, \emph{Maximum ratio transmission}, in \emph{1999 IEEE international
  conference on communications (Cat. No. 99CH36311)}, Vol.~2. IEEE, 1999, pp.
  1310--1314.

\bibitem{MMSE}
A.H. Mehana and A. Nosratinia, \emph{Diversity of mmse mimo receivers}, IEEE
  Transactions on information theory 58 (2012), pp. 6788--6805.

\bibitem{RZF19}
L.D. Nguyen, H.D. Tuan, T.Q. Duong, and H.V. Poor, \emph{Multi-user regularized
  zero-forcing beamforming}, IEEE Transactions on Signal Processing 67 (2019),
  pp. 2839--2853.

\bibitem{OptimalRegularization}
C.B. Peel, B.M. Hochwald, and A.L. Swindlehurst, \emph{A vector-perturbation
  technique for near-capacity multiantenna multiuser communication-part i:
  channel inversion and regularization}, IEEE Transactions on Communications 53
  (2005), pp. 195--202.

\bibitem{IRC}
B. Ren, Y. Wang, S. Sun, Y. Zhang, X. Dai, and K. Niu, \emph{Low-complexity
  mmse-irc algorithm for uplink massive mimo systems}, Electronics Letters 53
  (2017), pp. 972--974.

\bibitem{PrecodingDetection}
S. Shi, M. Schubert, and H. Boche, \emph{Downlink mmse transceiver optimization
  for multiuser mimo systems: Duality and sum-mse minimization}, IEEE
  Transactions on Signal Processing 55 (2007), pp. 5436--5446.

\bibitem{SVD}
L. Sun and M.R. McKay, \emph{Eigen-based transceivers for the mimo broadcast
  channel with semi-orthogonal user selection}, IEEE Transactions on Signal
  Processing 58 (2010), pp. 5246--5261.

\bibitem{SE}
S. Verd{\'u}, \emph{Spectral efficiency in the wideband regime}, IEEE
  Transactions on Information Theory 48 (2002), pp. 1319--1343.

\bibitem{wang2014sinr}
B. Wang, Y. Chang, and D. Yang, \emph{On the sinr in massive mimo networks with
  mmse receivers}, IEEE Communications Letters 18 (2014), pp. 1979--1982.

\bibitem{RZF}
Z. Wang and W. Chen, \emph{Regularized zero-forcing for multiantenna broadcast
  channels with user selection}, IEEE Wireless Communications Letters 1 (2012),
  pp. 129--132.

\bibitem{Inverses}
A. Wiesel, Y.C. Eldar, and S. Shamai, \emph{Zero-forcing precoding and
  generalized inverses}, IEEE Transactions on Signal Processing 56 (2008), pp.
  4409--4418.

\bibitem{MMSE2}
D. Wubben, R. Bohnke, V. Kuhn, and K.D. Kammeyer, \emph{Near-maximum-likelihood
  detection of MIMO systems using MMSE-based lattice-reduction}, in \emph{2004
  IEEE International Conference on Communications (IEEE Cat. No. 04CH37577)},
  Vol.~2. IEEE, 2004, pp. 798--802.

\bibitem{ZF}
T. Yoo and A. Goldsmith, \emph{On the optimality of multiantenna broadcast
  scheduling using zero-forcing beamforming}, IEEE Journal on selected areas in
  communications 24 (2006), pp. 528--541.

\bibitem{Per_antenna_const}
W. Yu and T. Lan, \emph{Transmitter optimization for the multi-antenna downlink
  with per-antenna power constraints}, IEEE Transactions on signal processing
  55 (2007), pp. 2646--2660.

\bibitem{LBFGS}
C. Zhu, R.H. Byrd, P. Lu, and J. Nocedal, \emph{Algorithm 778: L-bfgs-b:
  Fortran subroutines for large-scale bound-constrained optimization}, ACM
  Transactions on mathematical software (TOMS) 23 (1997), pp. 550--560.

\end{thebibliography}

\begin{table}
    \centering
    \begin{tabular}{|l|l|l|l|l|l|l|l|l|}
    \toprule
    Precoding &   MRT &     ZF &    RZF &   ARZF &  QN &  QN  &  QN  &  QN  \\
    \textrm{SUSINR} &       &        &        &         &   CD RZF         &    CD ARZF         &    IRC RZF         &     IRC ARZF         \\
    \midrule
    -4     &  0.60 &   0.26 &   0.59 &   1.58 &       1.92 &        1.92 &        2.13 &         2.10 \\
     0     &  0.79 &   0.52 &   0.85 &   2.25 &       2.70 &        2.69 &        2.95 &         2.95 \\
     4     &  0.98 &   0.95 &   1.24 &   3.02 &       3.59 &        3.64 &        3.91 &         3.94 \\
     8     &  1.16 &   1.56 &   1.78 &   3.89 &       4.59 &        4.70 &        4.99 &         5.00 \\
     12    &  1.33 &   2.38 &   2.52 &   4.86 &       5.73 &        5.85 &        6.17 &         6.20 \\
     16    &  1.51 &   3.39 &   3.46 &   5.90 &       6.92 &        7.11 &        7.42 &         7.47 \\
     20    &  1.70 &   4.57 &   4.61 &   7.04 &       8.17 &        8.42 &        8.71 &         8.81 \\
     24    &  1.91 &   5.93 &   5.95 &   8.29 &       9.43 &        9.83 &       10.01 &        10.25 \\
     28    &  2.11 &   7.41 &   7.42 &   9.59 &      10.67 &       11.28 &       11.25 &        11.68 \\
     32    &  2.29 &   9.15 &   9.15 &  11.06 &      11.99 &       12.80 &       12.57 &        13.18 \\
     36    &  2.45 &  11.06 &  11.06 &  12.63 &      13.53 &       14.35 &       13.97 &        14.64 \\
     40    &  2.56 &  13.14 &  13.14 &  14.31 &      15.28 &       15.90 &       15.45 &        16.08 \\
    \bottomrule
    \end{tabular}
    \caption{Urban NLOS 8 Users. The table shows IRC Spectral Efficiency of the different precoding algorithms.}
    \label{Urban NLOS 8 Users SE-IRC}
\end{table}

\begin{table}
    \centering
    \begin{tabular}{c | c}
        Parameter & Value \\
        \hline
        Base station parameters &  \\
        \hline
        number of base stations & $1$ \\
        position, m: (x, y, z) axes & $(0, 0, 25)$ \\
        number of antenna placeholders (y axis) & $8$ \\
        number of antenna placeholders (z axis) & $4$ \\
        distance between placeholders ($y$ axis) & $0.5$ wavelength \\
        distance between placeholders ($z$ axis) & $1.7$ wavelength \\
        antenna model & 3gpp-macro \\
        half-Power in azimuth direction, deg & 60 \\
        half-Power in elevation direction, deg & 10  \\
        front-to back ratio, dB  & 20 \\
        total number of antennas & 64 \\
        \hline
         Receiver parameters &  \\
        \hline
        number of placeholders at the receiver ($x$ axis) & $2$ \\
        distance between placeholders ($x$ axis) & $0.5$ wavelength \\
        antenna model & half-wave-dipole \\
        total number of antennas & 4 \\
        \hline
        Quadriga simulation parameters &  \\
        \hline
        central band frequency & $3.5$ GHz \\
        1 sample per meter (default value) & 1 \\
        include delay of the LOS path & 1 \\
        disable spherical waves (use\_3GPP\_baseline) & 1 \\
        \hline
        Quadriga channel builders parameters &  \\
        \hline
        shadow fading sigma & 0 \\
        cluster splitting & False \\
        bandwidth & $100$ MHz \\
        number of subcarriers & 42 \\ \\
    \end{tabular} 
    \caption{Quadriga generation parameters.}
    \label{tab:quadriga_hyp}
\end{table}

\end{document}